# The effect of bilayer domains on electronic transport properties of epitaxial graphene on SiC

Tom Yager[a], Arseniy Lartsev[a], Rositza Yakimova[b], Samuel Lara-Avila[a] *, Sergey Kubatkin[a]

[a]Department of Microtechnology and Nanoscience, Chalmers University of Technology, Göteborg, SE-41296, Sweden
[b]Department of Physics, Chemistry and Biology, Linköping University, Linköping, SE-58183, Sweden

Magnetotransport measurements on Hall bar devices fabricated on purely monolayer epitaxial graphene on Silicon Carbide (SiC/G) show a very tight spread in carrier concentration and mobility across wafer-size dimensions. In contrast, SiC/G devices containing bilayer graphene domains display variations in their electronic properties linked to the amount of bilayer content.

## 1. Introduction

Thermal decomposition of silicon carbide (SiC) at high temperatures is an attractive technological route to produce graphene on large areas, which is a practical requisite for the development of scalable graphene devices.[1,2] One of the main challenges in the technology of epitaxial graphene on SiC (SiC/G) is to produce electrically homogeneous monolayer graphene on a wafer scale. Since the electronic properties of SiC/G are related to its morphology, complete structural homogeneity of monolayer growth has been pursued by controlling the initial morphology of SiC,[3] minimizing the substrate mis-orientation,[1,4] employing different SiC polytypes,[2,5] and in general, by optimizing growth conditions.[1,2,6–8] However, growth kinetics of graphene on SiC invariably leads to formation and uncontrolled reconstruction of steps at the substrate surface and successive formation of terraces.[9,10] In principle, graphene layers nucleate at the step edges and once monolayer grows on the SiC surface, no more Si can sublimate leading to a self-limiting growth process.[7,11] Nevertheless, in practice, new graphene domains can still grow at the SiC-graphene interface even if the surface is already covered with graphene.[12,13] Not surprisingly, when the electronic properties of SiC/G are probed on the large scale (e.g. mm's size) variability of ~1-2 orders of magnitude in electronic properties such as resistivity, carrier density and mobility have been observed.[14] It is thus of practical importance to understand the origin of variations of the electronic transport properties of large scale SiC/G and reveal the impact of microscopic imperfections such as steps and bilayer domains.

*Corresponding Author. Email: samuel.lara@chalmers.se

In this work we report the effect of bilayer graphene domains on the electronic transport properties of predominantly monolayer SiC/G. By carefully aligning micron scale Hall bar devices on a well-characterised SiC/G wafer it is possible to gain new insight on the correlation of bilayer coverage and substrate steps on the electrical properties of SiC/G. We show that devices patterned on a macroscopic (7x7 mm$^2$) SiC/G wafer show no appreciable deviation in carrier concentration and mobility, provided they are free of multilayer graphene domains. On the other hand, the spread of electrical properties observed for SiC/G devices containing bilayer domains is directly linked to the amount of bilayer contained in each device. In agreement with previous reports,[15] the role of unit cell-high steps and their orientation relative to Hall bar devices do not lead to a wide variation in electrical properties.

## 2. Experimental

We fabricated Hall bar devices on SiC/G grown at high temperature and high pressure environment on the Si-face of 4H-SiC. The high temperature sublimation technique allows a reduction of the presence of high step edges, due to a smoothening of the surface, and thus reducing the formation of long bilayer bands to micron scale domains.[10] Prior to device patterning, the as-grown SiC/G substrate was mapped using optical microscopy in order to identify regions where monolayer or bilayer was predominant. Using optical microscopy it is possible to distinguish monolayer from bilayer graphene on SiC since the optical transparency of graphene on SiC is ~2.4% and ~3.7% for monolayer and bilayer, respectively.[15] From optical characterization, the SiC/G chip was found to have a continuous electrical monolayer of graphene as well as the presence, in places, of micron scale bilayer features. Further mapping at higher resolution was carried out by atomic force microscope (AFM), which requires extremely clean graphene surfaces. These are available on pristine samples after growth, but challenging to recover after fabrication due to the presence of nanometre scale polymer resist residuals. For this reason, gold alignment markers were directly evaporated on SiC/G through an etched silicon shadow mask onto the pristine surface prior to patterning, thus avoiding the use of polymer resists. This allowed twelve 80 μm x 80 μm areas to be scanned by high resolution height and phase AFM on marked regions to determine the local SiC/G step and layer properties (Fig. 1a). [16,17] The advantage of using phase-contrast AFM to identify graphene domains on the substrate lies on that it yields a bi-modal distribution of phase intensities for monolayer and bilayer, enabling reliable, high resolution quantification of the bilayer-to-monolayer ratio for each device.[16] Following substrate mapping, micron scale Hall bar devices were patterned on domains containing pre-determined proportions of monolayer and bilayer, with orientations selected to be perpendicular or parallel to the step orientation. In total 25 Hall bar devices with dimensions $W = 2$ μm x $L = 8$ μm for monolayer and $W = 1$ μm x $L = 4$ μm for bilayer were patterned on SiC/G using standard electron-beam lithography, lift-off, and oxygen plasma etching, as reported elsewhere.[5] The amount of bilayer present was determined independently for each measured region of the device, ranging from 0% to approximately 97%. Finally, before performing magnetotransport measurements, the sample was encapsulated with a resist bilayer in order to prevent doping drift upon thermal cycling and/or exposure to ambient conditions.[18] The electronic properties of the SiC/G reported here were obtained via magnetotransport measurements at room and cryogenic temperature in magnetic fields up to $B = 14$ T.

**3. Results and Discussion**

Fig. 1a shows a typical phase-contrast AFM image obtained on the surface of pristine SiC/G in which bilayer domains appear as dark regions on otherwise monolayer graphene (light contrast). The insets of Fig. 1a show two regions where devices have been patterned on either entirely monolayer regions or with a given amount of bilayer domains. Fig. 1b shows a phase-contrast AFM image of a Hall bar containing approximately 97% coverage of bilayer domains, estimated directly from the AFM scans. Throughout the entire chip, Hall bar devices were patterned with orientations along steps and across steps on SiC, which have a typical height < 2nm as obtained from height AFM measurements.

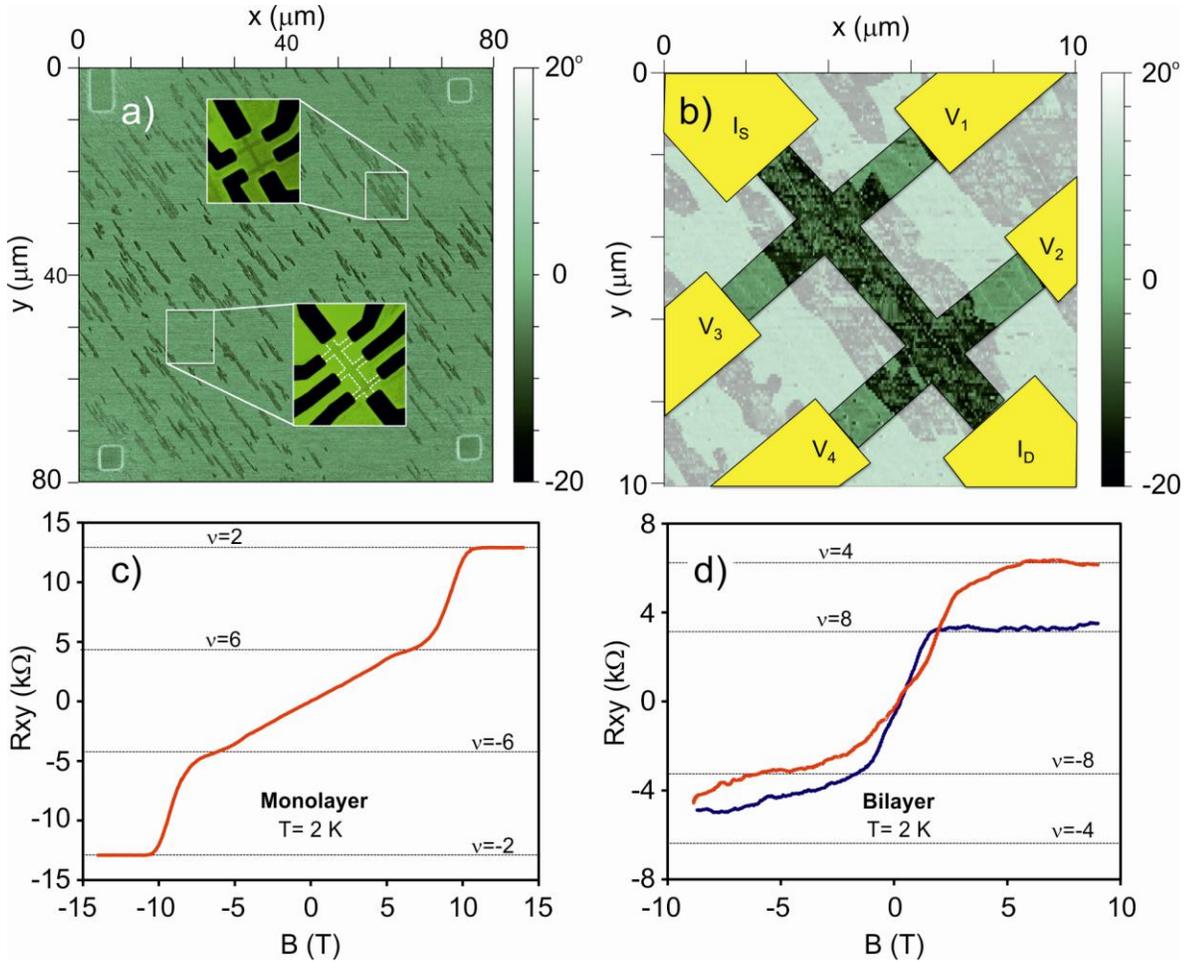

Fig. 1.

Hall bar devices with monolayer graphene and with a pre-determined amount of bilayer graphene coverage. a) Phase-contrast AFM imaging on a clean SiC/G surface (80 x 80 mm$^2$) provide a local map of the SiC/G and enable further alignment and micro-fabrication of Hall bar devices containing a given amount of bilayer graphene, from 0 to 97%. Shadow-evaporated gold markers (rectangles in the corners) were used to give alignment accuracy better than 300 nm. Insets show representative optical images of one monolayer and one bilayer device. b) Phase-contrast AFM image of a hall bar device containing 97% of bilayer domains. Bilayer domains appear as dark areas c) Plateaux at filling factors $\nu = \pm 2, \pm 6$ are a manifestation of half integer quantum Hall effect and confirm the monolayer nature of graphene. d) Quasi-plateaux at filling factors $\nu = 4, \pm 8$ in bilayer devices suggests that transport is indeed dominated by bilayer graphene.

The mono- and bi-layer nature of the graphene domains in each Hall bar was further investigated by low temperature magnetotransport at high magnetic fields, since the number of layers that effectively contribute to electron transport can be revealed by quantum-Hall effects (QHE's). For example, in monolayer graphene the quasirelativistic nature of carriers is manifested in an anomalous set of Hall plateaus at well-defined values of the transversal resistance, $R_{xy} = (h/4e^2)/(n+1/2) \approx 12.9, 4.3, 2.6…k\Omega$, with $h$ the Planck constant, $e$ the elementary charge, and $n$ integer, zero included. In contrast, for bilayer graphene with massive carriers, the plateaux sequence changes to $R_{xy} = h/(4e^2 n) \approx 6.5, 3.2, 2.1…k\Omega$ ($n \geq 1$). A practical challenge in "counting" layers in SiC/G samples by low temperature magnetotransport is that low carrier concentrations are needed to observe QHE's at experimentally available magnetic fields. To mention, when SiC/G samples are covered with resist the observed carrier density, obtained by low-field Hall measurements as $n = 1/eR_H = 1/e(dR_{xy}/dB)$, are of the order of ~ 2-3 $\times 10^{12}$ cm$^{-2}$ for nominally monolayer graphene devices, and the corresponding carrier concentration for devices containing up to 97% of bilayer coverage is of the order of $n = 1/eR_H \sim 8 \times 10^{12}$ cm$^{-2}$. These doping levels would require unfeasible magnetic fields of the order of ~ 80 T, unattainable in our set-up, limited to $B = 14$ T.

The observation of QHE's in the gated SiC/G devices, shown in Fig. 1c,d, confirmed their mono- and bi-layer nature as inferred from optical and AFM characterization. In order to decrease the carrier concentration in our devices to values suitable for observing QHE's we have used photochemical and corona ion gating.[18,19] With a carrier concentration in nominally monolayer devices reduced to $n \sim 7 \times 10^{11}$ cm$^{-2}$ by photochemical gating (doping range for this method limited to $\Delta n \sim 1\text{-}2 \times 10^{12}$ cm$^{-2}$), the observation of half-integer QHE with onset of plateaux $R_{xy} = h/(2e^2\nu)$ at filling factors $\nu = \pm 2, \pm 6$ confirmed the monolayer nature of the graphene layer in these devices (Fig. 1c). For bilayer graphene devices with an initial concentration $n = 1/eR_H \sim 8 \times 10^{12}$ cm$^{-2}$ corona ions were used to decrease carrier density to $n \sim 10^{11}$ cm$^{-2}$. At this low carrier density it was possible to observe traces of unconventional quantum Hall effect with quasi-plateau at filling factors $\nu = 4, \pm 8$, particular to bilayer graphene.[20] Fig. 1d shows low temperature magnetotransport for a bilayer sample at two different carrier density values attained with corona ion gating. Imperfection of the observed quantum Hall effect in bilayer devices may arise from the structural inhomogeneity of bilayer domains grown on this sample,[21–23] and the opening of a transport gap in bilayer at high electric field.[24–26]

The properties of 25 polymer-encapsulated monolayer and bilayer devices were assessed at room temperature prior to photochemical or corona ion gating. We start the discussion with 14 purely monolayer devices spread over the surface of a 7 x 7 mm$^2$ wafer. Significantly, these monolayer devices showed no notable difference in electron doping (Fig. 2a) or Hall mobility, even when oriented either parallel (9 devices) or perpendicular (5 devices) to silicon carbide steps on the substrate (Fig. 2b). The electron doping in these monolayer devices was found to be $n = (2.73 \pm 0.05) \times 10^{12}$ cm$^{-2}$ and the corresponding sheet resistivity $\rho_{xx} = R_{xx}W/L = 1.74 \pm 0.08$ k$\Omega$ sq$^{-1}$. (Fig. 2b). In contrast to the relatively tight spread of electrical properties observed in purely monolayer devices, Hall bar devices containing different proportions of bilayer graphene were found to have a wide variation of electrical properties depending on the amount of bilayer coverage. For all 25 devices placed millimetres apart on the chip with varying bilayer/monolayer ratio $\chi$ the electron density was found to vary between $n = (2.73 \pm 0.05) \times 10^{12}$ cm$^{-2}$ for $\chi = 0$ (purely monolayer) and $n = (7.6 \pm 0.6) \times 10^{12}$ cm$^{-2}$ for $\chi = 0.95$-$0.97$. In terms of resistivity, this was found to change from $\rho_{xx} = 1.74 \pm 0.08$ k$\Omega$ sq$^{-1}$ to $0.97 \pm 0.05$ k$\Omega$ sq$^{-1}$ for the same values of $\chi$. The variation of both carrier concentration and resistivity is seemingly linearly dependent on the bilayer coverage ratio $\chi$ within the range covered in this experiment.

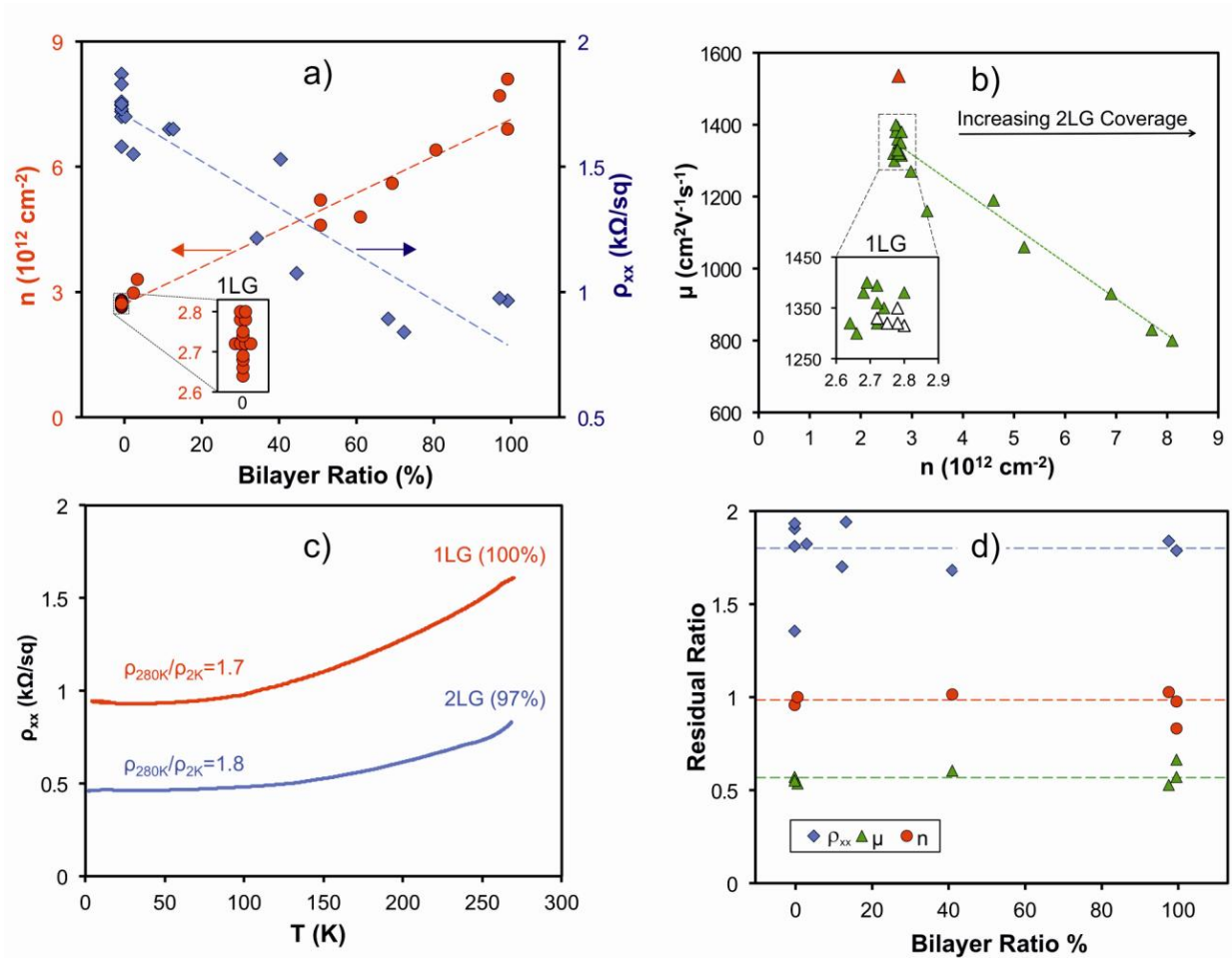

Fig. 2.

Magnetotransport characterization of 25 polymer-encapsulated Hall bar crosses. a) Electron doping (resistivity) increases (decreases) linearly for increasing proportion of bilayer graphene in the device. Note the tight spread of concentration for monolayer (inset). b) Carrier mobility as a function of carrier concentration at room temperature. Concentration was found to depend on the bilayer ratio. Inset: tight spread of carrier mobility for monolayer graphene, containing both perpendicular (white) and parallel (red) surface steps for 14 Hall bar crosses spread over the surface of a 7 x 7 mm$^2$ wafer. c) Comparison of the temperature dependence of resistance for a homogeneously monolayer and bilayer device. d) Temperature residual ratio for resistivity (blue diamonds), electron concentration (red circles) and mobility (green triangles) are independent of bilayer coverage from T=280 K to T=2 K.

The mechanism whereby bilayer graphene domains display higher electron doping can be understood by considering the combined effect of substrate and resist doping on the graphene layers. It has been shown from angle-resolved photoelectron spectroscopy (ARPES) that as-grown samples of epitaxial graphene on the Si-face of SiC are heavily electron doped. This has been explained, from electrostatic considerations, to be a consequence of workfunction difference between graphene and SiC as well as donor states on the interface (buffer) layer. ARPES experiments show that the Fermi level lies ~0.42 eV above the Dirac point for monolayer graphene,[27] and for bilayer on SiC, gapped due to substrate-induced asymmetry between the layers ($E_{g-2L}$ ~ 0.1 eV), the Fermi level lies ~ 0.4 eV above the middle of the gap.[25,27] In terms of electron density, this corresponds to $n_{1L} = 4\pi E_F^2/h^2 v_F^2 \approx 1.2$x $10^{13}$ cm$^{-2}$ and $n_{2L} = 8\pi E_F^2/h^2 v_F^2$ in the range 2.4x $10^{13}$ cm$^{-2}$.[28] When encapsulated by resist, the carrier density in monolayer and bilayer SiC/G are both decreased, and the difference in magnitude of the carrier density shift is attributed to the difference in responsivity to an electrostatic gate in these materials.[28,29]

The Hall mobility, estimated as $\mu = \rho_{xx}/R_H$ was found to be $\mu = 1350 \pm 30$ cm$^2$ V$^{-1}$ s$^{-1}$ for 14 monolayer devices and $\mu = 850 \pm 70$ cm$^2$ V$^{-1}$ s$^{-1}$ for 3 devices with 95-97% coverage of bilayer (Fig. 2b). When all devices are taken into account the carrier mobility decreases linearly as function of carrier concentration, which in our samples is related to bilayer coverage. The source of mobility degradation in the presence of bilayer domains may be explained by the presence of resistive dipoles at the monolayer-bilayer interface.[30] Thus, for homogeneously monolayer devices the spread in carrier concentration and mobility are minimal. In fact, the outlier observed at $\mu = 1550$ cm$^2$V$^{-1}$s$^{-1}$, $n = 2.7 \times 10^{12}$ cm$^{-2}$ in Fig. 2b (red triangle) is an important demonstration that the properties of the Hall cross and Hall channel cannot always be assumed to be equal. The Hall resistance ($R_H$) characterizes the material within the Hall cross of the device, whereas longitudinal resistivity ($\rho_{xx}$) provides assessment of resistance within the Hall channel. In this case, the Hall cross in the device was 100% monolayer whereas the Hall channel contained a significant proportion of bilayer. Combining the measurements at each region without taking into account the local presence of bilayer domains (such as in the case of carrier mobility $\mu = \rho_{xx}/R_H$), will result in the erroneous perception of spread in electronic properties in monolayer SiC/G across a chip.

The temperature dependence has also been studied for selected devices down to $T = 2$ K, and the residual ratio of transport properties was not found to depend significantly on the amount of bilayer domains in the device. For example, Fig. 2c shows the resistivity of an entirely monolayer sample and a nearly bilayer sample as function of temperature. The residual resistance ratio $RRR = \rho_{270K}/\rho_{2K}$ is 1.7 and 1.8 respectively for monolayer and bilayer devices, attributed to suppression of acoustic phonons at low temperatures.[31,32] Fig. 1d display that the residual ratios calculated for resistivity, carrier concentration and electron mobility are essentially independent of bilayer coverage, and show that while the electron density remains essentially constant, the mobility increased by nearly a factor of 2 at low temperatures.

## 4. Conclusions

In summary, we find that when devices are carefully fabricated on SiC/G, the spread in device-to-device electronic properties (carrier concentration and mobility) is substantially reduced provided the devices are fabricated entirely on monolayer domains. The spread in properties among devices patterned on the same SiC/G wafer can thus be understood by considering the inhomogeneous number of layers often grown on the surface of epitaxial graphene on SiC. We confirmed the substrate steps of height < 2 nm, which can be achieved by high temperature sublimation technique, do not have a profound impact on the electrical properties of SiC/G. We therefore conclude that monolayer SiC/G is actually homogeneous at the wafer scale and identify bilayer domains as the source of the apparent spread in SiC/G electronic transport properties observed in previous reports. Our findings highlight the importance of achieving homogeneous graphene layers on SiC by eliminating the presence of multilayer domains in order to enable large-scale integration of SiC/G-based devices. Additionally, we highlight the importance of carefully patterning devices on purely monolayer regions whenever the properties of SiC/G are investigated.

**Acknowledgements**

This work was partly supported by The Graphene Flagship (Contract No. CNECT-ICT-604391), Swedish Foundation for Strategic Research (SSF), Linnaeus Centre for Quantum Engineering, Knut and Allice Wallenberg Foundation, Chalmers AoA Nano, and the EMRP project GraphOhm. The EMRP is jointly funded by the EMRP participating countries within EURAMET and the European Union.